\documentstyle[aps]{revtex}

\newcommand{\cplus}{\dot+}
\newcommand{\ts}{\left(}
\newcommand{\td}{\right)}
\newcommand{\qs}{\left[}

\newcommand{\qd}{\right]}

\newcommand{\nn}{\nonumber}
\newcommand{\x}{{\bf x }}
\newcommand{\mA}{\mathcal{A}}
\newcommand{\mN}{\mathcal{N}}
\newcommand{\mP}{\mathcal{P}}
\newcommand{\mL}{\mathcal{L}}
\newcommand{\de}{{\mathrm{d}}}
\newcommand{\kkP}{$\kappa$-Poincar\'{e}}
\newcommand{\kM}{$\kappa$-Minkowski}

\newcommand{\cM}[1]{[M_j,{#1}_k]=i\epsilon_{jkl}{#1}_l}
\newcommand{\Da}[1]{\Delta(#1)=#1\otimes 1+1\otimes #1}

\def\be{\begin{equation}}
\def\ee{\end{equation}}
\def\bea{\begin{eqnarray}}
\def\eea{\end{eqnarray}}
\newcommand{\e}{\epsilon}

\newcommand{\ar}{P_0,P^2}

\begin{document}

\title{\begin{flushright}
hep-th/0306013 \\
$~$
\end{flushright}
{\Large {\bf
Hopf-algebra description of noncommutative-spacetime symmetries}}
}

\author{$~$\\
{\bf Alessandra~AGOSTINI}$^a$, {\bf Giovanni~AMELINO-CAMELIA}$^b$
and {\bf Francesco~D'ANDREA}$^b$}
\address{$^a$Dipartimento di Scienze Fisiche, Universit\`{a} di Napoli ``Federico II'',
\\
Monte S.~Angelo, Via Cintia, 80126 Napoli, Italy
\\
$^b$Dipartimento di Fisica,
Universit\`{a} di Roma ``La Sapienza'' and INFN Sez.~Roma1,\\
P.le Moro 2, 00185 Roma, Italy}

\maketitle

\begin{abstract}\noindent%
In the study of certain noncommutative versions
of Minkowski spacetime there is still
a large ambiguity concerning the characterization
of their symmetries.
Adopting as our case study the \kM\ noncommutative space-time,
on which a large literature is already available,
we propose a line of analysis of noncommutative-spacetime symmetries
that relies on the introduction of a  Weyl map (connecting
a given function in the noncommutative Minkowski  with a
corresponding function in commutative Minkowski)
and of a compatible notion of integration in the noncommutative spacetime.
We confirm (and we establish more robustly) previous suggestions that
the commutative-spacetime notion of Lie-algebra symmetries must be replaced,
in the noncommutative-spacetime context, by the
one of Hopf-algebra symmetries.
We prove that in \kM\ it is possible to construct an action which is
invariant under
a Poincar\'{e}-like Hopf algebra of symmetries with 10 generators,
in which the noncommutativity length scale has the role of relativistic
invariant.
The approach here adopted does leave one residual ambiguity,
which pertains to the
description of the translation generators, but our results,
independently of this ambiguity, are sufficient to clarify
that some recent studies (gr-qc/0212128 and hep-th/0301061),
which argued for an
operational indistiguishability between theories with and without
a length-scale relativistic invariant, implicitly assumed that
the underlying spacetime would be classical.
\end{abstract}



\bigskip
\bigskip



\date{\today}


\section{Introduction}\noindent
Non-commutative versions of Minkowski
spacetime~\cite{books,Maj00-Fou}
are being considered in several quantum-gravity scenarios.
In some cases
non-commutative spacetimes prove
useful at an effective-theory level (for example, in certain string theory
pictures~\cite{SW,susskind,douglasnovikov} spacetime noncommutativity
provides an effective theory description of
the physics of strings in presence of a corresponding external background),
while other quantum-gravity approaches~\cite{Snyder,DFR,dsr1,Kow02-NST}
explore the possibility
that a noncommutativity might be needed for the correct
fundamental description of spacetime (and a noncommutative version
of Minkowski spacetime would be relevant in the flat-spacetime
limit).

In recent research on these noncommutative spacetimes much attention
has been devoted to the implications of noncommutativity
for the classical
Poincar\'{e} symmetries of Minkowski spacetime.
For the simplest noncommutative versions of Minkowski spacetime,
the canonical noncommutative spacetimes characterized by coordinate
noncommutativity of type ($\mu=0,1,2,3$)
\bea
&&[{\x}_\mu,{\x}_\nu]= i\theta_{\mu,\nu}\label{canonical}
\eea
with coordinate-independent $\theta_{\mu,\nu}$,
a full understanding has been matured, and in particular it has been
established that the Lorentz-sector symmetries are broken
(see, {\it e.g.}, Refs.~\cite{susskind,douglasnovikov,gacluisa})
by this type of noncommutativity.

Unfortunately, already at the next level of complexity, the one of
Lie-algebra noncommutative spacetimes
\begin{equation}\label{liealg}
[{\x}_\mu,{\x}_\nu]= i \zeta_{\mu,\nu}^{\sigma} {\x}_\sigma ~,
\end{equation}
our present understanding of the fate of Poincar\'{e} symmetries
is still unsatisfactory. For some of these Lie-algebra
spacetimes there has been much discussion
in the mathematical literature
(see, {\it e.g.}, Refs.~\cite{lukieAnnPhys,Kow02-NST})
about their ``duality", in a certain mathematically-defined sense,
to corresponding Hopf-algebra
versions of the Poincar\'{e} algebra. Moreover, in
studies exploring the recent proposal of
physical theories with two relativistic
invariants~\cite{dsr1,Kow02-NST,Mag01-DSR}
it has been conjectured
that Lie-algebra noncommutative
spacetimes might play a role in the new class
of relativistic theories (see, {\it e.g.}, Refs.~\cite{dsr1,Kow02-NST}).
However, a full characterization of the fate of
Poincar\'{e} symmetries in Lie-algebra
noncommutative spacetimes is still missing.

Here we intend to contribute to the understanding of these delicate
issues.
We focus on the illustrative example
of the much studied ``\kM\ Lie-algebra noncommutative
spacetime"\cite{MajidRuegg,lukieAnnPhys},
whose coordinates
satisfy the commutation relations
\be
[{\x}_j,{\x}_0]=i\lambda {\x}_j~,~~~[{\x}_j,{\x}_k]=0
\label{eq:kM}
\ee
where $j,k=1,2,3$ and we denote the noncommutativity
parameter by $\lambda$ (in some other \kM\ studies the reader will
encounter the parameter $\kappa$, which is related to the $\lambda$
of (\ref{eq:kM}) by  $\lambda=\kappa^{-1}$).

It is puzzling that in the literature it is sometimes stated
(see, {\it e.g.}, Refs.~\cite{lukieAnnPhys,Kow02-NST})
that the symmetries of \kM\ can be described
by any one of a large number
of \kkP\ Hopf algebras. The nature of this claimed
symmetry-description degeneracy
remains obscure from a physics perspective, since it is only
supported by an (equally obscure) ``duality"
criterion~\cite{lukieAnnPhys,Kow02-NST}.

This issue has recently taken central stage also in the mentioned
research on the physical proposal~\cite{dsr1} of relativistic theories
with two invariants, the so-called DSR theories,
where \kM\ is being considered as a possible example of spacetime
that hosts two relativistic invariants.
In that context only the Lorentz sector is considered, but again
(since different \kkP\ algebras lead to different types of
Lorentz transformations) the symmetry-description degeneracy
is posing a key challenge~\cite{Kow02-NST}.
In some related studies~\cite{ahluOPERAT,grumi}
it was even argued that the presence of two
relativistic invariants should be seen as a mathematical/formal
artifact, {\it i.e.} the relativistic theory with two invariants
(independently of a possible role for Hopf algebras) could be
equivalently described as a relativistic theory with Lie-algebra symmetries
and a single invariant.
A corollary of our results will prove that the line of
analysis advocated in Ref.~\cite{ahluOPERAT,grumi} implicitly
assumes that the underlying spacetime be classical.
In certain quantum spacetimes, {\it e.g.} in some noncommutative spacetimes,
the presence of the length (or energy) scale as relativistic invariant
is unavoidable and reflects genuine physical aspects of the theory.

A key point that emerges from the study we report in the following
sections is that in certain noncommutative spacetimes
(in particular in \kM) different physical theories formulated in
that same spacetime might enjoy different types of 10-generator
Poincar\'{e}-like symmetries.
Whereas theories in classical (commutative) Minkowski spacetime
can only be Poincar\'{e} invariant, if they enjoy the maximal number
of symmetries, we expect that different theories in \kM\ spacetime
can enjoy different types of 10-generators symmetries\footnote{Of course,
both in Minkowski and in \kM\ one can also have theories with less
than 10 symmetries, symmetry-breaking theories,
but this will not be our focus here.}.

In this first study adopting our new approach,
the concept of ``theory in \kM\ spacetime" is simply identified with
the introduction of an action.
We emphasize the role that a Weyl map and a compatible notion of integration
can play in characterizing the action and its symmetries.
We call \emph{symmetry operator} an operator over \kM\
that leaves invariant the action of the theory.
We also stress that the symmetries may
also depend on the structure of the differential calculus
introduced in the noncommutative spacetime (since the differential
calculus affects the formulation of the action).

We start by reviewing some of the main properties of \kM\ noncommutative
spacetime and of Weyl maps that are naturally considered in the
analysis of \kM\ and by reviewing (and providing arguments in favour of)
the use of Hopf algebras in the description of the symmetries
of a noncommutative spacetime.
Then in the second part of the paper we apply these concepts in the
analysis of the action for a free scalar field in \kM\ spacetime.

\section{\kM\ Space-time}\noindent
The elements of \kM~\cite{MajidRuegg} spacetime are sums and products
of the elements ${\x}_\mu$
(the ``noncommutative coordinates'') which
satisfy the commutation relations $[{\x}_j,{\x}_k]=0$, $[{\x}_j,{\x}_0]
=i\lambda {\x}_j$,
which we already noted in Eq.~(\ref{eq:kM}).
[\kM\ is the universal enveloping algebra of the Lie algebra (\ref{eq:kM}).]
Of course,
conventional commuting coordinates are recovered in the limit $\lambda\to 0$,
and it is possible to establish a one-to-one correspondence between elements
of \kM\ and analytical functions of four commuting variables $x_\mu$.
Such a correspondence is called \emph{Weyl map}, and is not unique.
Since one of the points we want to emphasize is the possible dependence
of the symmetry analysis on the choice of the Weyl map it is convenient
for us to focus on two specific choices, which we denote by $\Omega_R$
and $\Omega_S$. It is sufficient to specify the Weyl map on
the complex exponentials
and extend it to the generic function $f(x)$, whose Fourier transform is
\mbox{$\tilde{f}(p)=\frac{1}{(2\pi)^4}\int f(x)e^{-ipx}{\de}^4p$},
by linearity
\begin{displaymath}
\Omega_{R,S}(f)=\int \, \tilde{f}(p)
\, \Omega_{R,S}(e^{ipx}) \, {\de}^4p
~.
\end{displaymath}

Our first choice of Weyl map is the ``time-to-the-right" map, which
is most commonly adopted
in the \kM\ literature (see, {\it e.g.}, Ref.~\cite{AmelinoMajid}):
\begin{eqnarray}
\Omega_R(e^{ipx}) = e^{i\vec{p}\vec{\x}}e^{-ip_0\x_0} \label{right}
\end{eqnarray}
and the alternative we also consider is the ``time-symmetrized"
map~\cite{ALZ}
\begin{eqnarray}
\Omega_S(e^{ipx}) = e^{-ip_0{\x}_0/2}e^{i\vec{p}\vec{\x}}e^{-ip_0{\x}_0/2}
\label{sym}
\end{eqnarray}
[We are adopting conventions such
that $px=\vec{p}\vec{x}-p_0x_0$, with $p_\mu$ four real
commuting parameters, and
similarly $p{\x}=\vec{p}\vec{\x}-p_0 {\x}_0$.]

It is straightforward to verify that these Weyl maps are equivalent to
corresponding ordering conventions.
The $\Omega_R$ map associates to each commutative function the
element obtained ordering the time $x_0$ to the right and replacing the $x$
variables with the elements ${\x}$. For example
\begin{eqnarray*}
\Omega_R(x_0x_j) = \Omega_R(x_jx_0)={\x}_j{\x}_0
\end{eqnarray*}
(and notice that ${\x}_j{\x}_0 \neq {\x}_0{\x}_j = {\x}_j({\x}_0-i\lambda)
=\Omega_R(x_j x_0-i\lambda x_j)$).
Naturally one can refer to this choice as
\emph{time--to--the--right ordering} or \emph{right ordering}
convention. The $\Omega_S$ map, instead, associates to each commutative
function a time-symmetrized element, for example
\begin{displaymath}
\Omega_S(x_0x_j)=\Omega_S(x_jx_0)=\frac{1}{2}({\x}_0{\x}_j+{\x}_j{\x}_0)
\end{displaymath}
and could be described
as \emph{time-symmetrized ordering} or \emph{symmetric ordering}.

For the benefit of readers more familiar with the use of the
star-product in noncommutative spacetimes we stress that
the choice of a Weyl map is also equivalent to the choice of a star-product
representation of \kM. The star-products associated to $\Omega_R$ and
$\Omega_S$ are respectivedly defined by
\begin{eqnarray*}
\Omega_{R}(f\,*_R\,g) &=& \Omega_{R}(f){\cdot}\Omega_{R}(g)\\
\Omega_{S}(f\,*_S\,g) &=& \Omega_{S}(f){\cdot}\Omega_{S}(g)
\end{eqnarray*}
The space of functions equipped with one of the products $*_R$/$*_S$ is
a noncommutative algebra isomorphic to \kM\ (each star-product is a different
representation of the same algebra).

While in canonical spacetimes, (\ref{canonical}),
the star-product formulas are characterized by a standard addition
of momenta (Fourier paramenters), in Lie-algebra spacetimes the
form of the commutators among coordinates imposes a more complicated
formulation of the composition of Fourier parameters.
In the case of \kM, and of our specific choices of Weyl maps,
one finds
\begin{eqnarray*}
\Omega_R(e^{ipx}){\cdot} \Omega_R(e^{iqx})
&=& \Omega_R(e^{i(\vec{p}+\vec{q}e^{-\lambda p_0})\vec{x}-i(p_0+q_0)x_0}) \\
\Omega_S(e^{ipx}){\cdot} \Omega_S(e^{iqx})
&=& \Omega_S(e^{i(\vec{p}e^{\frac{\lambda}{2}q_0}
+\vec{q}e^{-\frac{\lambda}{2}p_0})\vec{x}-i(p_0+q_0)x_0})
\end{eqnarray*}
The exponentials form the \emph{unitary group} of the algebra, and the last
equations furnish the multiplication law in two different representations of
the group.

In closing this section we also note a relation between the two
Weyl maps here considered
\be
\Omega_R(e^{ipx})=\Omega_S(e^{i\vec{p}e^{\frac{\lambda}{2} p_0}\vec{x}-ip_0x_0})
\ee
{\it i.e.} it is possible to go from time-to-the-right to
time-symmetrized ordering through a four-momenta transformation.
This will also play a key role in our analysis.

\section{Hopf-Algebra description of symmetries}\noindent
In this work we consider only \emph{external symmetries}.
In the familiar context of theories in commutative spacetimes
we describe an external symmetry as a
transformation of the coordinates that
leaves invariant the action of the theory,
and we shall insist on this property in the case
of noncommutative spacetimes.

In preparation for our analysis it is useful to consider
the simple action $S(\phi)=\int{\de}^4x\;\phi(\Box - M^2)\phi$
of a free scalar
field $\phi$ in commutative Minkowski spacetime
($\Box=\partial_\mu\partial^\mu$ is the familiar D'Alembert operator).
We will briefly review the Lie-algebra description of the
symmetries of this action and then show that, upon considering
appropriate coalgebraic properties, this can be
cast in Hopf-algebra language. However, for theories in commutative
spacetime the coalgebra aspects of the symmetry transformations
are inevitably trivial, essentially contained already in the statement
of the algebra aspects of the symmetry transformations, and
it is therefore appropriate to specify the symmetries exclusively
in Lie-algebra language, as commonly done.
We will then argue that for theories in noncommutative spacetimes
the coalgebraic properties are not in general trivial and a full
characterization of the symmetries requires the full language
of Hopf algebras.

So we start with our free scalar field in commutative Minkowski spacetime,
and we observe that the most general
infinitesimal transformation that can
be considered is $x'_\mu=x_\mu+\epsilon A_\mu(x)$,
with $A_\mu$ four real functions of the coordinates.

For the scalar field $\phi'(x')=\phi(x)$ and in leading order in $\epsilon$
one finds
\begin{displaymath}
\phi'(x)-\phi(x)=\{\partial^\mu\phi(x)\}(x_\mu-x'_\mu)=-\epsilon
A_\mu(x)\partial^\mu\phi(x)
\end{displaymath}
In terms of the generator $T$ of the
transformation, $T=iA_\mu(x)\partial^\mu$,
one obtains $x'=(1-i\epsilon T)x$ and $\phi'=(1+i\epsilon T)\phi$.
[The action of $T$ over $\x$ is indicated by $T\x$.]

The variation of the action, at the leading order in $\epsilon$ is
\bea
S(\phi')-S(\phi)&=&i\epsilon\!\int\!{\de}^4x\left(T\{\phi
(\Box-M^2)\phi\}+\phi[\Box,T]\phi\right)\nn\\
&=&i\epsilon\!\int\!{\de}^4x\left(T{\mL}(x)+\phi[\Box,T]\phi\right)\nn
\eea
and therefore the action is invariant under $T$-generated transformations
if and only if
\bea
\int\!{\de}^4x\left(T{\mL}(x)+\phi[\Box,T]\phi\right) = 0
~.\label{twoparts}
\eea

Let us pause to briefly comment on the structure of
the condition (\ref{twoparts}).
In the case of the illustrative example we are here considering
of a free scalar field in commutative Minkowski spacetime
there is a well-established form of the (maximally-symmetric)
action and one can just verify that condition (\ref{twoparts})
is satisfied.
In cases in which the (possibly noncommutative) spacetime is given
and one is looking for a maximally-symmetic form of the action
that analysis can naturally progress in two steps: in the
first step one
can determine the algebra whose elements $T$ satisfy
\begin{equation}\label{eq:sym}
\int{\de}^4x\;T\,{\mL}(x)=0
\end{equation}
for every\footnote{Indipendently of
the form of the differential operator ${\cal {O}}$ contained in ${\mL}(x)$
the condition that
$\int{\de}^4x\;T\,\{\phi {\cal {O}} \phi)\}=0$
for a generic $\phi$
is equivalent to the condition
$\int{\de}^4x\;T\,{\mL}(x) = 0$ for a generic ${\mL}$.}
scalar function ${\mL}(x)$
and then in the second step one can construct the
Lagrangian ${\mL}(x)$ of the theory in terms of $\phi(x)$ and
of a differential operator ${\cal {O}}$ imposing $[{\cal {O}},T]=0$
(in our illustrative example ${\cal {O}} = \Box-M^2$).

This observation will prove useful as we later look for a
maximally-symmetric action in $\kappa$-Minkowski.
For the case of a free scalar field in commutative Minkowski spacetime
the choice  ${\cal {O}} = \Box-M^2$ is well established and
leads to a maximally-symmetric action.
The symmetries of this action are described in terms of
the classical Poincar\'{e} algebra $\mP$, generated
by the elements
\begin{displaymath}
P_\mu=-i\partial_\mu\quad M_j=\epsilon_{jkl}x_kP_l\quad N_j=x_jP_0-x_0P_j
\end{displaymath}
which satisfy the commutation relations
\begin{eqnarray*}
& [P_\mu,P_\nu]=0\qquad [M_j,P_0]=0\qquad \cM{P} & \\
& \cM{M}\qquad\cM{N} & \\ & [N_j,P_0]=iP_j \quad [N_j,P_k]=i\delta_{jk}P_0
\quad [N_j,N_k]=-i\epsilon_{jkl}M_l &
\end{eqnarray*}
The operator $\Box=-P_\mu P^\mu$ is the first casimir of the algebra,
and of course satisfies $[\Box,T]=0$.

For this case of a maximally-symmetric theory
in commutative Minkowski spacetime
it is conventional to describe the symmetries fully in terms
of the Poincar\'{e} Lie algebra.
For \kM\ noncommutative spacetime we shall argue that a description
in terms of a Hopf algebra is necessary.
In order to guide the reader toward this step we want to first
show that even in the commutative-Minkowski case there is an
underlying Hopf-algebra structure, but the commutativity of
functions in Minkowski spacetimes implies that the additional
structures present in the Hopf algebra (with respect to the Lie algebra)
are all trivial. We will then observe that the noncommutativity of
functions in noncommutative \kM\ spacetime leads to a nontrivial
Hopf-algebra structure, which cannot be faithfully captured in the simpler
language of Lie algebras.

In preparation for this exercise it is convenient to introduce
a map $\epsilon$ such that
\begin{equation}\label{eq:Usym}
\int{\de}^4x\;U\,{\mL}(x)=\,\epsilon(U)\int{\de}^4x\;{\mL}(x)
\end{equation}
for each element $U\in\mP$ and for each function ${\mL}(x)$. It is
straightforward to verify that $\epsilon(1)=1$ (with an abuse of notation, we
indicate with $1$ the identity transformation) and
of course from (\ref{eq:sym}) it follows that $\epsilon(T)=0$ for each
generator $T$ of the algebra.
Moreover, from (\ref{eq:Usym}) it follows that
\begin{displaymath}
\int\!{\de}^4x\,U(U'{\mL})=\epsilon(U)\!\!\int\!{\de}^4x\,U'{\mL}=
\epsilon(U)\epsilon(U')\!\!\int\!{\de}^4x\,{\mL}
\end{displaymath}
But the first member is also equal to $\epsilon(UU')\!\int\!{\de}^4x\,{\mL}$,
and therefore $\epsilon$ is an algebra morphism
\begin{displaymath}
\epsilon(UU')=\epsilon(U)\epsilon(U')
\end{displaymath}
This recursive formula allows us to
calculate $\epsilon$ for a generic element of the algebra.
[Also note that the application of $U$ to a constant
function $f=1$ gives $U{\cdot}1=\epsilon(U)$.]

For each $U\in\mP$ we can also introduce
a map $\Delta\!:\mP\to\mP\otimes\mP$ such
that for every $f(x)$ and $g(x)$
\begin{equation}\label{eq:cop}
U(f{\cdot} g)= (U_{(1)}f)(U_{(2)}g)
\end{equation}
where $\Delta(U)=\sum_i U_{(1)}^i\otimes U_{(2)}^i$
is written simply as $U_{(1)}\otimes U_{(2)}$ (Sweedler notation).
The reader can easily verify that $\Delta(1)=1\otimes 1$, and $\Da{T}$ for each
generator of Poincar\'{e} algebra. This last property, which plays a key role
in allowing a description of the symmetries at the simple Lie-algebra
level (without any true need to resort to a full Hopf-algebra description)
is actually connected with the commutativity of function in Minkowski spacetime.
In fact, from $f {\cdot} g = g {\cdot} f$, one easily finds that $\Delta$ is symmetric
\begin{displaymath}
 U_{(1)}\otimes U_{(2)}= U_{(2)}\otimes U_{(1)}
\end{displaymath}
for all $U$ or, adopting math gergeon, $\Delta$ is \emph{cocommutative}
(in a certain sense, $\Delta$ is trivial).
In general in a noncommutative spacetime $\Delta$ is not cocommutative.

Another noteworthy property,
the fact that $\Delta$ is coassociative, {\it i.e.} $ U_{(1)}\otimes\Delta
(U_{(2)})=\Delta (U_{(1)})\otimes U_{(2)}$,
follows straightforwardly from the
associativity of the product of functions.
From the property $(UU'){\cdot}(fg)=U(U'{\cdot} fg)$ it follows that $\Delta$
is an algebra morphism, {\it i.e.} $\Delta(UU')=\Delta(U)\Delta(U')$,
which allows to calculate $\Delta$ recursively.
And finally by considering (\ref{eq:cop}) for $g=1$ one
obtains $Uf=U(f{\cdot} 1)= (U_{(1)}f)(U_{(2)}1)= (U_{(1)}f)\epsilon(U_{(2)})$
from which it follows that $U_{(1)}\epsilon(U_{(2)})=U$ (and similarly
$\epsilon(U_{(1)})U_{(2)}=U$).

These two maps $\epsilon$ and $\Delta$,
defined by (\ref{eq:Usym}) and (\ref{eq:cop}),
make a generic symmetry algebra into
a \emph{bialgebra}. $\epsilon$ is called \emph{counit}
and $\Delta$ is called \emph{coproduct}.

Defining $S(1)=1$, $S(T)=-T$ for each generator and
$S(UU')=S(U')S(U)$ we obtain a map
satisfying $U_{(1)}S(U_{(2)})= S(U_{(1)})U_{(2)}=\epsilon(U)$.
This makes $\mP$ a Hopf algebra~\cite{Maj00-Fou}.
The map $S$ is called antipode
and (if it exists, as in this case) it is unique.

In an appropriate sense a Lie algebra is equivalent to
a ``trivial Hopf algebra", a Hopf algebra with the
trivial structure of counit,
coproduct and antipode which we just described.
For theories in commutative spacetime
the symmetries can always be described in terms of a trivial
Hopf algebra. In contemplating theories in noncommutative spacetime
it is natural to insist on the requirement that the symmetries
be described by a Hopf algebra. The Lie-algebra description cannot
be maintained, since it would not provide a sufficient set of rules
to handle consistently the laws of symmetry transformation
of products of (noncommutative) functions.
The requirement that symmetries be described in terms of a Hopf
algebra actually is a simple statement: the action of symmetry
transformations on products
of functions should be consistent with the fact that such
products are themself functions, and, accordingly,
the laws of transformation of products of functions should
still only require the appropriate action of the generators
of the (Hopf) algebra.

Once the algebra properties are specified (action of symmetry
transformations on functions of the noncommutative coordinates)
the properties of the counit, coproduct and antipode can
always be formally derived, but these will not in general satisfy
the Hopf algebra criteria since they may require the introduction
of new operators, not included in the algebra sector.
If this does not occur (if the counit, coprodut and antipode
that one obtains on the basis of the algebra sector can be
expressed fully in terms of operators in the algebra)
the Hopf-algebra criteria are automatically satisfied.

\section{Symmetries of a theory in \kM\ spacetime}
\subsection{General strategy}
In this first exploratory study in which our description of
symmetries is applied we only consider a free scalar theory
in \kM\ spacetime. This is a significant limitation. In fact, our
concept of symmetries applies directly to theories and not
to the underlying spacetime on which the theories are introduced,
so one may expect different results for different theories
(even restricting our attention to theories
10-generator  Poincar\'{e}-like symmetries).

Having specified our objective as the one of describing free scalar
fields in \kM\ we have not yet established the form of the action.
We will look for an action which realizes our objective of
a theory with symmetries described
by a 10-generator Hopf algebra.

As announced, our construction of the theory will for convenience
make use of Weyl maps, which allow us to keep track a various
properties in the familiar context of functions of auxiliary commutative
coordinates.
We have already introduced the Weyl maps $\Omega_R$ and $\Omega_S$
which shall be useful to discuss our results.
We are therefore ready for the first step in constructing
the action for a free scalar particle. Of course we need
a rule of integration.
In the \kM\ literature there is already substantial work
on rules of integration that are naturally expressed using
the $\Omega_R$ Weyl map
\begin{equation}
\int_R {\de}^4{\x}\;\Omega_R(f)=\int f(x)\,{\de}^4x \;\;.
\label{intR}
\end{equation}
Our alternative choice of Weyl map would naturally invite us to
consider the integration rule
\begin{equation}
\int_S {\de}^4{\x}\; \Omega_S(f)=\int f(x)\,{\de}^4x\;\;.
\label{intS}
\end{equation}
Actually these integrals are
equivalent, {\it i.e.} $\int_R{\de}^4{\x}\;\Phi=\int_S{\de}^4{\x}\;\Phi$
for each element $\Phi$ of \kM. This is easily verified by expressing
the most general element of \kM\ both in its $\Omega_R$-inspired
form and its $\Omega_S$-inspired form
\begin{eqnarray}
\Phi = \int{\de}^4p\;\tilde{f}(p)\Omega_R(e^{ipx}) =
\int{\de}^4p\;\tilde{f}(p_0,pe^{-\lambda p_0/2})e^{-3\lambda
p_0/2}\Omega_S(e^{ipx})
\label{rightandsymm}
\end{eqnarray}
and observing that
\begin{displaymath}
\int_R{\de}^4{\x}\;\Phi=\int_S{\de}^4{\x}\;\Phi = (2\pi)^4\tilde{f}(0)
~.
\end{displaymath}
In our search of a maximally-symmetric theory with construction
based on $\Omega_R$ or $\Omega_S$ we therefore have a natural candidate
for the integration rules to be used: (\ref{intR}),
which can be equivalently reformulated as (\ref{intS}).
[Because of the equivalence we will omit indices $R$ or $S$ on
the integration symbol.]

We can now also start formulating an educated guess for the general
structure of the action we are seeking
\begin{displaymath}
S(\Phi)=\int{\de}^4{\x}\;\Phi(\Box_\lambda-M^2)\Phi
\end{displaymath}
where $\Phi$ is a generic real\footnote{Of course, there is
no intuitive concept of ``reality" for a function of noncommuting coordinates;
however, it is natural to state that $\Phi$ is a real element of \kM\
if it is obtained from a real commutative function through the $\Omega$-map.}
element of \kM, $M^2$ is  (real, dimensionful and) positive
and $\Box_\lambda$ is a (differential) operator which is
still to be specified (we need more guidance concerning
our requirement of obtaining a maximally-symmetric theory)
but we know it should reproduce the familiar D'Alembert operator
in the $\lambda \rightarrow 0$ commutative-spacetime limit.
For each real element $\Phi$, the action $S(\Phi)$ is a real number.

By straightforward generalization of the results
reviewed in the previous section, we pose
that a transformation $T$ will be a symmetry if (and only if)
\be
\int{\de}^4{\x}\;\left(T{\cdot}\left\{\Phi\left(\Box_\lambda-M^2\right)\Phi\right\}+
\Phi[\Box_\lambda,T]\Phi\right)=0 \label{twopartsnc}~.
\ee
When several such symmetry generators are available they may or may
not combine together to form a Hopf algebra.
When they do form a Hopf algebra, denote it generically by $\mA$,
and we attribute to $\mA$ the role of symmetry (Hopf) algebra.

As mentioned in the preceding section, the search of a maximally-symmetric
action can be structured in two steps.
In the first step one looks for a Hopf algebra (in our case a Hopf algebra
which has the Poincar\'{e} algebra as
classical limit) whose generators $T$ satisfy
\begin{equation}\label{eq:kMsym}
\int{\de}^4{\x}\;T{\mL}({\x})=0
\end{equation}
for each element ${\mL}({\x})$ of \kM.
In the second step one looks for an operator $\Box_\lambda$
that is invariant ($[\Box_\lambda,T]=0$) under the action of this algebra.

\subsection{Translations}
In introducing the concept of translations we of course want
to follow as closely as possible the analogy with the well-established
concepts that apply in the commutative limit $\lambda \rightarrow 0$.
Since we have defined functions in \kM\ in terms of the Weyl maps,
and since the Weyl maps are fully specified once given on
Fourier exponentials (then the map on generic functions simply
requires the introduction of standard Fourier transforms),
we can, when convenient, confine the discussion to the Fourier
exponentials.
And an ambiguity, which is deeply connected with noncommutativity,
confronts us immediately:
in commutative Minkowski
the translation generator acts according to
\bea
P_{\mu}(e^{ikx}) &=& k_{\mu} e^{ikx}
\label{eq1}
\eea
but this action of the translation generators cannot
be implemented on general functions of the \kM\ coordinates.
The root of the ambiguity can be exposed
by just considering the same function of  \kM\ coordinates
written in two ways, the time-to-the-right form
and the time-symmetrized form.
Let us first write a specific function $\Phi$ of the \kM\ coordinates
in the way suggested by the time-to-the-right Weyl map:
\begin{eqnarray}
\Phi = \int\de^4p\;\tilde{f}(p)\Omega_R(e^{ipx})
~.
\label{eq2}
\end{eqnarray}
On the basis of Eqs.~(\ref{eq1}) and (\ref{eq2})
it would seem natural to define translations in \kM\ as
generated by the operators $P^R_{\mu}$ such that
\bea
P^R_{\mu} \Omega_R(e^{ikx}) = k_{\mu} \Omega_R(e^{ikx})
~.
\label{eq3}
\eea
But, as already stated through Eq.~(\ref{rightandsymm}),
the same function $\Phi$ written in Eq.~(\ref{eq2})
using time-to-the-right ordering can also be equivalently
expressed in time-symmetrized form as
\begin{eqnarray}
\Phi = \int\de^4p\;\tilde{f}(p_0,pe^{-\lambda p_0/2})
e^{-3\lambda p_0/2}\Omega_S(e^{ipx})
\label{eq4}
\end{eqnarray}
and on the basis of Eqs.~(\ref{eq1}) and (\ref{eq4})
it would seem natural to define translations in \kM\ as
generated by the operators $P^S_{\mu}$ such that
\bea
P^S_{\mu} \Omega_S(e^{ikx}) = k_{\mu} \Omega_S(e^{ikx})
~.
\label{eq5}
\eea

We had already encountered an ``ordering ambiguity"
in introducing a law of integration in \kM, but there we eventually
realized that there was no ambiguity after all (the two approaches
to the law of integration led to identical results).
The ordering ambiguity we are facing now in defining translations
is certainly more serious. In fact, the two candidates as
translation generators $P^S_{\mu}$ and $P^R_{\mu}$
are truly inequivalent, as the careful reader can easily
verify by applying $P^S_{\mu}$ and $P^R_{\mu}$ to
a few examples of functions in \kM; in particular:
\begin{eqnarray}
P^R_{\mu} (e^{i\vec{k}\vec{\x}}e^{-ik_0{\x}_0})
 & \! = \! &
P^R_{\mu} \Omega_R(e^{ikx}) =
k_{\mu} \Omega_R(e^{ikx}) =
k_{\mu} (e^{i\vec{k}\vec{\x}}e^{-ik_0{\x}_0})=
k_{\mu} ( e^{-ik_0{\x}_0/2}e^{
ie^{\frac{\lambda}{2} k_0}\vec{k}\vec{\x}}e^{-ik_0{\x}_0/2}) \nn\\
& \! \neq \! & e^{\frac{\lambda}{2} k_0} k_{\mu}
( e^{-ik_0{\x_0}/2}e^{
ie^{\frac{\lambda}{2} k_0}\vec{k}\vec{\x}}e^{-ik_0{\x}_0/2})
= P^S_{\mu} (e^{i\vec{k}\vec{\x}}e^{-ik_0{\x}_0})
~.
\label{eq6}
\end{eqnarray}

It is also easy to verify that both $P^S_{\mu}$ and $P^R_{\mu}$
satisfy condition (\ref{eq:kMsym}):
\begin{equation}
\int{\de}^4{\x}\; P^{R,S}_{\mu} {\mL}({\x})=0
\end{equation}
Moreover the quadruplet of operators $P^S_{\mu}$ and
the quadruplet of operators $P^R_{\mu}$ do separately
give rise to genuine Hopf algebras of translation-like
symmetry transformations.
Since, as mentioned, the exponentials $e^{i\vec{k}\vec{\x}}e^{-ik_0\x_0}$
form a basis of \kM,
the coproduct of the $P^R_{j}$ operators, $\Delta P_j^R$,
is obtained consistently from observing that
\bea
&&P^R_j\Omega_R(e^{ikx})\Omega_R(e^{ipx})
=-i\Omega_R(\partial_je^{i(k\cplus p)x})\nn\\
&&=-i\Omega_R((k\cplus_R p)_je^{i(k\cplus p)x})\nn\\
&&=[P^R_j\Omega_R(e^{ikx})][\Omega_R(e^{ipx})]
+[e^{-\lambda P^R_0}\Omega_R(e^{ikx})][P^R_j\Omega_R(e^{ipx})]
~,
\label{calccoprod}
\eea
where $p \cplus q \equiv (p_0+q_0, p_1 + q_1 e^{-\lambda p_0},
 p_2 + q_2 e^{-\lambda p_0}, p_3 + q_3 e^{-\lambda p_0})$,
which implies
\be
\Delta P^R_j=P_j^R\otimes 1
+ e^{-\lambda P_0^R}\otimes P^R_j\label{coprodpr}
\ee
Following an analogous procedure one can derive
\begin{eqnarray*}
& \Da{P_0^R} &
\end{eqnarray*}
and the full structure of a four-generator Hopf algebra emerges.

The same goes through, with equal success, for the $P^S_{\mu}$
alternative.
The coproducts naturally take a different form,
\bea
\Delta P^S_0&=&P^S_0\otimes 1+1\otimes P^S_0\nn\\
\Delta P^S_j&=&P_j^S\otimes e^{\frac{\lambda}{2}P_0^S}
+ e^{-\frac{\lambda}{2}P_0^S}\otimes P^S_j
\label{coprodps}
\eea
but the full structure of a four-generator Hopf algebra
is again straightforwardly obtained.

We must therefore live with this ambiguity. As we look
for an example of maximally-symmetric theory in \kM\
both the option $P^R_{\mu}$ and the option $P^S_{\mu}$ must be
considered\footnote{Of course, it is reasonable to contemplate
even more options in addition to these two; however, for our purposes
it is convenient to focus on these two examples. This will be sufficient
in order to illustrate the differences and the common features
of the two choices (similar differences and analogous common features
should be expected of other possibilities which might be considered).
We also stress that the two options we consider, the one inspired by
time-to-the-right ordering and the one inspired by time-symmetrized
ordering, can be viewed as the most natural ones. In fact,
the time coordinate has a privileged role in the \kM\ commutation
relations and it is natural to think first of all to three
options: time-symmetrized, time-to-the-right and time-to-the-left.
We are actually considering also the time-to-the-left option; in fact,
as the careful reader can easily verify,
the implications of time-to-the-left ordering are indistinguishable
from the ones of the time-to-the-right ordering (identical Hopf-algebra
structures...just a few trivial formal differences).}, {\it i.e.}
we can look for 10-generator extensions of either.

\subsection{Rotations}
Next we attempt to obtain a 7-generator Hopf algebra, describing
four translation-like operators and three rotation-like generators.
Since, as just mentioned, our analysis of translations alone
led us to two alternatives, we are in principle prepared
for at least two alternative version of the 7-generator
Hopf-algebra for translations and rotations:
($P^R,M^R$) and ($P^S,M^S$), with $M^R$ and $M^S$ to be determined.

For what concerns the translations we have found that an
acceptable Hopf-algebra description was obtained by
straightforward ``quantization"
of the classical translations:
the $P^R_{\mu}$ translations where just obtained from the
commutative-spacetime translations through the $\Omega_R$
Weyl map and the $P^S_{\mu}$
translations where just obtained from the
commutative-spacetime translations through the $\Omega_S$
Weyl map.

Also noticing that the \kM\ commutation relations are structured in
a way that appears to be very respectful of classical rotations,
it is natural to explore the possibility of describing also the
rotations by straightforward ``quantization"
\begin{eqnarray}
 && M^R_j{\cdot}\Omega_R(f) = \Omega_R (M_j f)
 =\Omega_R(-i\epsilon_{jkl}x_k\partial_l f) \label{rotar}\\
 && M^S_j{\cdot}\Omega_S(f) = \Omega_S (M_j f)
 =\Omega_S(-i\epsilon_{jkl}x_k\partial_l f)
 \label{rotas}
\end{eqnarray}
where again in defining these rotation generators
we used the fact that any element of \kM\ (any ``function
in \kM\ spacetime") can be obtained through the
action of a Weyl map from some commutative
function $f(x)$.

This is actually another instance in which an ordering ambiguity
is only apparent. In fact, $M_j^S$ and $M_j^R$ act exactly in
the same way, as one can verify through the observation that
\bea
M_j^S\Omega_R(e^{ikx})&=&M_j^S\Omega_S(e^{ik'x})_{k'=(k_0,k_ie^{\lambda k_0/2})}\nn\\
=-i\epsilon_{jkl}\Omega_S(x_k\partial_le^{ik'x})&=&\epsilon_{jkl}\Omega_S(x_ke^{\lambda k_0/2}k_le^{ik'x})\nn\\
=-i\epsilon_{jkl}k_l\Omega_S(\partial_{k_k}e^{ik'x})&=&-i\epsilon_{jkl}k_l\partial_{k_k}\Omega_S(e^{ik'x})\nn\\
=-i\epsilon_{jkl}k_l\partial_{k_k}\Omega_R(e^{ikx})&=&\epsilon_{jkl}k_l\Omega_R(-i\partial_{k_k}e^{ikx})\nn\\
=-i\epsilon_{jkl}\Omega_R(x_k\partial_le^{ikx})&=&M_j^R\Omega_R(e^{ikx})\nn
\eea
We will therefore remove the indices $R$/$S$ on the rotation generator,
and denote simply by $M_j$.

One easily verifies that the operators $M_j$, as defined by (\ref{rotar})
(and equivalently defined by (\ref{rotas})),
are good candidates for
the construction of (Hopf) algebras of
symmetries of a theory in \kM\ spacetime; in fact:
\begin{equation}
\int{\de}^4{\x}\; M_j {\mL}({\x})=0 ~.
\end{equation}

It is also straightforward to verify that
\bea
[M_j,M_k]=i\varepsilon_{jkl} M_l
\label{rotalgebra}
\eea
and, following a line of analysis analogous to the one
of Eq.~(\ref{calccoprod}),
that a trivial coproduct must be adopted for
these candidate rotation operators:
\bea
\Delta M_j=M_j\otimes 1+1\otimes M_j
\label{rotcoprod}
\eea
Therefore the triplet $M_j$
forms a 3-generator Hopf algebra
that is completely undeformed (classical) both in the algebra and
in the coalgebra sectors. (Using the intuitive description introduced
earlier this is a trivial rotation Hopf algebras, whose structure could
be equally well  captured by the standard Lie algebra of rotations.)

There is therefore a difference between the translations sector
and the rotations sector. Both translations and rotations can
be realized as straightforward (up to ordering) quantization
of their classical actions, but while for rotations even the
coalgebraic properties are classical (trivial coalgebra)
for the translations we found
a nontrivial coalgebra sector (\ref{coprodpr})
(or alternatively (\ref{coprodps})).

Up to this point we have two candidates, $P^S_\mu$
and $P^S_\mu$, for a  4-generator
Hopf algebra of translation symmetries and a candidate, $M_j$,
for a 3-generator
Hopf algebra of rotation symmetries. These can actually be
put together straightforwardly to obtain two candidates
for 7-generator translations-rotations symmetries ($P^R_\mu,M_j$)
and ($P^S_\mu,M_j$).
It is sufficient to observe that
\be
\left.
\begin{array}{ll}
[M_j,P^{R,S}_{\mu}]{\cdot} \Omega(e^{ikx})
=\varepsilon_{jkl}\Omega([-x_k\partial_{\mu}
+\partial_{\mu}x_k]\partial_l e^{ikx})&\\
=\delta_{\mu k}\varepsilon_{jkl}\Omega(\partial_le^{ikx})&
\end{array}
\right.
\ee
from which it follows that
\be
[M_i,P^{R,S}_j]=i\varepsilon_{ijk}P^{R,S}_k,\;\;
[M_i,P^{R,S}_0]=0
\ee
{\it i.e.} the action of rotations on energy-momentum
is undeformed. Accordingly, the generators $M_j$
can be represented as differential operators over
energy-momentum space in the familiar
way, $M_j=-i\varepsilon_{jkl}P_k\partial_{P_l}$.
[This also provides another opportunity
to verify (\ref{rotalgebra}).]

\subsection{Failure of classical Boosts}
In the analysis of translations and boosts
we have already encountered two different situations: rotations
in \kM\ spacetime are essentially classical in all respects,
while translations
have a ``classical" action
(straightforward $\Omega$-map ``quantization" of the
corresponding classical action)
but have nontrivial coalgebraic properties.
As we now intend to include also boosts, and obtain 10-generator
symmetry algebras, we encounter another possibility:
for boosts not only the coalgebra sector is nontrivial but
even the action on functions in \kM\ cannot be obtained
by ``quantization" of the classical action.

Because of its profound implications for our analysis,
we devote this subsection to the analysis of classical boosts,
showing that they do not lead to acceptable symmetries.
For simplicity we only look for
boosts $N_j^R$
that could combine with the translations $P^R_\mu$ and rotations $M_j$
to give us a 10-generator symmetry algebra.
The assumption that the action of these boosts could be ``classical"
means
\bea
N_j^R\Omega_R(f)=\Omega_R(N_jf)
=\Omega_R(i[x_0\partial_j-x_j\partial_0]f])
~.
\eea
It is easy to see (and it is obvious) that these boosts
combine with the rotations $M_j$ to close  a healthy
undeformed Lorentz algebra, and that adding also the
translations $P_\mu^R$ one obtains
the undeformed Poincar\'{e} algebra.
However, these algebras cannot be extended (by introducing
a suitable coalgebra sector) to obtain a (Hopf) algebra of symmetries
of theories in our noncommutative \kM\ spacetime.
In particular, we find an inconsistency in the coproduct of
these boosts $N_j^R$, which signals an obstruction originating
from an inadequacy in the description of the action
of boosts on (noncommutative) products of \kM\ functions.

The coproduct of the boosts $N_j^R$ can be inferred from observing that
\bea
&&N_j^R \Omega_R(e^{ikx})\Omega_R(e^{ipx})=\Omega_R(N_je^{i(k\cplus p)x})\nn\\
&&=-\Omega_R[(x_0(k\cplus p)_j-x_j(k+p)_0)e^{i(k\cplus p)x}]\nn\\
&&=-[\Omega_R([k_jx_0-k_0x_j]e^{ikx})\Omega_R(e^{ipx})\nn\\
&&+\Omega_R(e^{-\lambda k_0}e^{ikx})\Omega_R([p_jx_0-p_0x_j]e^{ipx})\nn\\
&&-\lambda\Omega_R(k_je^{ikx})\Omega_R(p_lx_le^{ipx})
+2\Omega_R(\sinh(\lambda k_0)e^{ikx})\Omega_R(p_0x_je^{ipx})]\nn
\eea
where we also used the relations
\bea
\Omega_R(x_0e^{i(k\cplus p)x})&\!
=\!&\Omega_R(e^{i(k\cplus p)x}){\x}_0\nn\\
=\Omega_R(e^{ikx}){\x}_0 \Omega_R(e^{ipx})&\!-\!&\lambda p \Omega_R(e^{ikx}){\x}\Omega_R(e^{ipx})\nn\\
=\Omega_R(x_0e^{ikx}) \Omega_R(e^{ipx})&\!-\!&\lambda \Omega_R(e^{ikx})\Omega_R(p_lx_le^{ipx})\nn\\
\Omega_R(x_j e^{i(k\cplus p){\x}})&=&{\x}_j \Omega_R(e^{i(k\cplus p)x})\nn\\
=e^{\lambda k_0}\Omega_R(e^{ikx}){\x}_j \Omega_R(e^{ipx})&\!=\!&e^{\lambda k_0}\Omega_R(e^{ikx})\Omega_R(x_je^{ipx})\nn
\eea
 From this one concludes that the coproduct is
\begin{eqnarray}
\Delta(N_j^R) &=& N_j^R\otimes 1+e^{-\lambda P_0^R}\otimes N_j^R+\nn\\ &&
+2\sinh\lambda P_0^R \otimes{\x}_jP_0^R-\lambda P_j^R\otimes\vec{\x}\vec{P}_R
\label{oldboostcoprod}
\end{eqnarray}
The problem is that $\Delta(N_j^R)$ is not an element of the algebraic tensor
product, {\it i.e.} it is not a function only of the elements $M,N,P$.
It would
be necessary to introduce new elements to close the Hopf algebra; for example,
the well known \emph{dilatation operator} ${\cal D}=\vec{\x}\vec{P}_R$, which
is independent from the other generators.
Therefore, as anticipated, the ``classical" choice $N_j^R$ cannot be combined
with $M^R$ and $P^R_\mu$ to obtain a 10-generator symmetry algebra.
From (\ref{oldboostcoprod}) one might immagine that perhaps an 11-generator
symmetry algebra, including the dilatation operator ${\cal D}$,
could be contemplated but this is also not possible since one can
easily verify that ${\cal D}$ does not satisfy condition (\ref{eq:kMsym}),
\begin{equation}
\int{\de}^4{\x}\;  {\cal D} {\mL}({\x}) \neq 0 ~.
\end{equation}
${\cal D}$ cannot be a symmetry of a scalar action in \kM.

Since the ``classical" choice $N_j^R$ is inadequate we are left
with two possible outcomes: either there is no 10-generator
symmetry-algebra extension
of the 7-generator symmetry algebra ($P^R_\mu$,$M_j$)
or the 10-generator
symmetry-algebra extension exists but requires nonclassical boosts.
In the next Subsection we show that the latter is true.

As mentioned the careful reader can easily verify that all these points
we made about the ``classical" $N_j^R$ and the ($P^R_\mu$,$M_j$,$N^R_j$)
algebra also apply to the ``classical" $N_j^S$
and the ($P^S_\mu$,$M_j$,$N^S_j$) algebra.

\subsection{Deformed boosts succeed}\noindent
We now show that there is a nonclassical choice of boosts ${\mN}_j^R$
that leads to a satisfactory ($P^R_\mu$,$M_j$,${\mN}^R_j$) symmetry (Hopf)
algebra.
On the basis of the problem encountered in the previous subsection
it is clear that our key task is to find
an operator ${\mN}_j^R$ such that $\Delta({\mN}_j^R)$ is
an element of $\mA\otimes\mA$, where $\mA$ is the algebra generated
by $(P^R_\mu,M_j,{\mN}_j^R)$.
We will show that this is possible, and that our ${\mN}_j^R$
has the correct classical limit and even the classical Lorentz-subalgebra
relations can be preserved (the commutation relations
involving $M_j,{\mN}_j^R$ are undeformed\footnote{The fact that
the Lorentz-subalgebra relations are undeformed is actually not surprising.
It is easy to see, just on the basis of dimensional considerations,
that the Lorentz algebra cannot be deformed by a length parameter $\lambda$.
In order to  deform the Lorentz-sector commutation relations one
should renounce to the property that the Lorentz-sector
generators form a subalgebra (the Lorentz-sector commutators can only
be deformed if energy-momentum generators are allowed to enter
the commutation relations). Fortunately we find that this is not required.},
{\it i.e.} $\lambda$-independent, and therefore exactly as in the
commutative Minkowski spacetime case).


Let us start by observing that, as one can easily verify,
by imposing that the deformed boost generator ${\mathcal{N}}_j$
(although possibly having a nonclassical action)
transforms as a vector under rotation one finds
\bea
{\mathcal{N}}_j \Omega(\phi(x))
=\Omega\{[ix_0A(-i\partial_x)\partial_j
+\lambda^{-1}x_jB(-i\partial_x)
-\lambda x_lC(-i\partial_x) \partial_{l} \partial_{j}
-i\epsilon_{jkl} x_k D(-i\partial_x)\partial_{l}]\phi(x)\} \label{boost}
\eea
where $A,B,C,D$ are unknown functions of $P^R_\mu$
(in the classical limit $A=i,D=0$; moreover, as $\lambda \rightarrow 0$
one obtains the classical limit if $\lambda C \rightarrow 0$
and $B \rightarrow \lambda P_0$).

Imposing
\be
{\mN}^R_j[\Omega(e^{ikx})\Omega(e^{ipx})]
=[{\mN}^R_{(1),j}\Omega(e^{ikx})][{\mN}^R_{(2),j}\Omega(e^{ikp})]
\ee
one clearly obtains some constraints on the functions $A,B,C,D$
that must be satisfied in order to
obtain the desired result that the operator ${\mN}_j^R$
is such that $\Delta({\mN}_j^R)$ is
an element of $\mA\otimes\mA$.
These constraints are discussed and analyzed in the Appendix.
The final result is
\be
{\mN}_j^R\Omega_R(f)=\Omega_R([ix_0\partial_j
+x_j(\frac{1-e^{2i\lambda \partial_0}}{2\lambda}-\frac{\lambda}{2}\nabla^2)
-\lambda x_l\partial_l\partial_j]f)\label{boostRfin}
\ee

It is easy to verify that the Hopf algebra $(P^R_\mu,M_j,{\mN}_j^R)$
satisfies all the requirements for a candidate symmetry-algebra
for theories in \kM\ spacetime.

Analogously on finds that the deformed boost generators
\be
{\mN}_j^S\Omega_S(f)=\Omega_S([ix_0\partial_j
-x_j(\frac{\sinh(i\lambda\partial_0)}{\lambda}+\frac{\lambda}{2}\nabla^2)
+\frac{\lambda}{2}x_l\partial_l\partial_j]e^{i\frac{\lambda}{2}\partial_0}f)
\label{boostSfin2}
\ee
combine with the translations $P^S_\mu$ and the rotations $M_j$
to give a genuine symmetry Hopf algebra $(P^R_\mu,M_j,{\mN}_j^S)$.

Although introduced following different formulations (respectively
the action on right-ordered functions and the action on symmetrically-ordered
functions) ${\mN}_j^R$ and ${\mN}_j^S$ are actually identical.
This is easily seen by observing that
\be
{\mN}_j^S\Omega_R(f)=\Omega_R([ix_0\partial_j
+x_j(\frac{1-e^{2i\lambda \partial_0}}{2\lambda}
-\frac{\lambda}{2}\nabla^2)-\lambda x_l\partial_l\partial_j]f)
\label{boostSfin1}
\ee
and comparing with (\ref{boostRfin}). We therefore remove the indices $R$/$S$
and use the unified notation ${\mN}_j$.

In summary we have two candidate Hopf algebras of
 10-generator Poincar\'{e}-like symmetries for \kM: $(P^R_\mu,M_j,{\mN}_j)$
 and $(P^S_\mu,M_j,{\mN}_j)$.

\subsection{An action for the free scalar field in \kM}
We have completed what we had announced as ``step 1" of our analysis:
we have constructed explicitly
candidates (actually not one but two candidates)
of 10-generator symmetry Hopf algebras for theories in \kM\ spacetime,
whose generators, generically denoted by $T$, satisfy
\begin{equation}\label{eq:kMsymbis}
\int{\de}^4{\x}\;T{\mL}({\x})=0
\end{equation}
for every element ${\mL}({\x})$ of \kM\ ({\it i.e.} for every
function ${\mL}({\x})$ of the \kM\ coordinates).

We are finally ready for the second step: introducing
a differential operator $\Box_\lambda$,
whose classical $\lambda \rightarrow 0$ limit
is the familiar D'Alembert operator, and such that the action
\be
S(\Phi)=\int{\de}^4{\x}\;\Phi(\Box_\lambda-M^2)\Phi \label{actionfinal}
\ee
is invariant under one of the Hopf algebras
we have constructed $(P^R_\mu,M_j,{\mN}_j)$
and  $(P^S_\mu,M_j,{\mN}_j)$.
We therefore must verify that, for some choice
of $\Box_\lambda$, $[\Box_\lambda,T]=0$ for every $T$ in the Hopf algebra.

For the $(P^R_\mu,M_j,{\mN}_j)$ case, guided
by the intuition that $\Box_\lambda$ should be a scalar
with respect to $(P^R_\mu,M_j,{\mN}_j)$ transformations,
one is led to the proposal
\be
\Box_\lambda=\left(\frac{2}{\lambda}\sinh\frac{\lambda P_0^R}{2}\right)^2-
e^{\lambda P_0^2}P_R^2\label{boxfin}
\ee

In fact, it is easy to verify that with this choice
of $\Box_\lambda$ the action (\ref{actionfinal}) is invariant
under the $(P^R_\mu,M_j,{\mN}_j)$ transformations. Therefore,
we have finally managed to
construct an action describing free scalar fields in \kM\
that enjoyes 10-generator (Hopf-algebra)
symmetries $(P^R_\mu,M_j,{\mN}_j)$.

It may surprise the reader
that this same choice of differential operator $\Box_\lambda$
is also acceptable from the perspective of the alternative
$(P^S_\mu,M_j,{\mN}_j)$ symmetries.
It turns out that the $\Box_\lambda$ given
in (\ref{boxfin}) is also a
scalar with respect to the $(P^S_\mu,M_j,{\mN}_j)$ transformations
($[\Box_\lambda,T]=0$ for every $T$ in $(P^S_\mu,M_j,{\mN}_j)$)
and the action (\ref{actionfinal}) with $\Box_\lambda$ given
in (\ref{boxfin}) is invariant also under the
$(P^S_\mu,M_j,{\mN}_j)$ transformations.

We will provide some additional remarks on this ambiguity
in the next Section, but let us first clarify what we found.
The action (\ref{actionfinal}) with $\Box_\lambda$ given
in (\ref{boxfin}) is invariant both
under $(P^R_\mu,M_j,{\mN}_j)$ transformations
and under $(P^S_\mu,M_j,{\mN}_j)$ transformations.
It is of course not proper to state that
it is invariant under 14-generator
$(P^R_\mu,P^S_\mu,M_j,{\mN}_j)$ transformations
(this 14-generator algebra is not a Hopf algebra and its
generators are not truly independent), but rather we have
an ambiguity in the description of the symmetries
of the action we have constructed for a free scalar field
in \kM. As discussed we are here
formulating the ambiguity in terms of two Hopf algebras,
$(P^R_\mu,M_j,N_j)$ and $(P^S_\mu,M_j,N_j)$,
which are profoundly connected with two different choices
of ordering in \kM\ (in the sense codified in the Weyl maps $\Omega_R$
and $\Omega$). We anticipate that the ambiguity should actually be
between more than two options: if one considered other types of ordering
conventions (ignoring the fact that these other alternatives might
not be ``natural")
for functions in \kM\ one would find other candidates $P^*_\mu$ as translation
generators and, correspondingly, other candidate
10-generator Hopf algebras of Poincar\'{e}-like symmetries
for \kM\ of the type $(P^*_\mu,M_j,{\mN}_j)$.

\section{Reducing the ambiguity to the choice of
translation generators}
The observation reported in the previous Section clearly confronts
us with an important aspect of the noncommutative geometries
we are considering. One could consider assuming that the two Hopf
algebras $(P^R_\mu,M_j,N_j)$ and $(P^S_\mu,M_j,N_j)$ are actually
equivalent dscriptions of the symmetries of the theory, but
this would only postpone the question at the level of describing
energy-momentum in our theory. We are used to associate energy-momentum
with the translation generators and it is not conceivable that a given
operative definition of energy-momentum could be equivalently
described in the context of the $(P^R_\mu,M_j,N_j)$ and $(P^S_\mu,M_j,N_j)$
Hopf algebras. If the properties of $P^R_\mu$ describe the properties
of a certain operative definition of energy-momentum then for that
same operative definition of energy-momentum a description
in terms of $P^S_\mu$ will turn out to be
unacceptable. The difference would be easily established by using,
for example, the different dispersion relations that $P^R_\mu$ and $P^S_\mu$
satisfy (a meaningful physical property, which could, in particular,
have observable consequences in astrophysics~\cite{grbgacgactPRDgianlufran}
and in cosmology~\cite{jurekCOSMO,maguCosmoDispRel}).

There are clearly several paths one could follow in exploring this issue.
One hypothesis that deserves investigation in future studies is the one
that perhaps the theory we considered does not have enough structure
to give proper physical significance to energy-momentum.
In the study of gauge theories in canonical noncommutative spacetime
it has emerged that gauge transformations and spacetime transformations
are deeply connected. It therefore seems important to aim for the construction
of gauge theories in \kM\ spacetime, which might have a deep role
in clarifying the status of the energy-momentum observables.

While clearly our analysis leaves this question unanswered,
we have reduced and clarifed the ambiguity that one finds in
previous \kM\ literature concerning the description of symmetries.
As mentioned in the Introduction, in the literature there
has been discussion
(see, {\it e.g.}, Refs.~\cite{lukieAnnPhys,Kow02-NST})
of a large class of Hopf algebras as candidates
for the description of the symmetries of \kM.~These
candidates were identified using a Hopf-algebra-duality
criterion.
The commutation relations that characterize these different
Hopf algebras fulfilling the duality condition take a very different
form in the different algebras, just like very different
commutation relations characterize our algebras $(P^R_\mu,M_j,N_j)$
and $(P^S_\mu,M_j,N_j)$. In the previous literature
the differences between the algebras were perceived in very general
terms, so much so that, in terms of our notation, our two algebras
would have been distinguished as $(P^R_\mu,M^R_j,N^R_j)$
and $(P^S_\mu,M^S_j,N^S_j)$.
Instead our two algebras involve the same generators for rotations
and boosts. These generators have different commutation rules
with translations, but
only because the translations $P^R_\mu$ are genuinely different
from the translations $P^S_\mu$. When acting on the same entity
(in particular when acting on functions of the \kM\ coordinates)
the generators for rotations and boosts of the two algebras
act in exactly the same way, they are identical operators.

We have therefore clarified that the difference between alternative
10-generator algebras of symmetries merely reflect a different
concept of translations. As discussed in Subsection~IV.B the type
of spacetime noncommutativity we are considering does naturally invite
one to consider alternative descriptions of translations.

\section{Aside on the new \kkP\ Hopf algebra that emerged from
the analysis}
While the work here reported has as primary objective the one of
contributing to the understanding of physical theories in
certain types of noncommutative spacetimes, we should at least mention
in passing that in the analysis reported in the previous Section
we have stumbled upon an interesting new example of
the so-called \kkP\ Hopf algebras.
The reader familiar with the subject will recognize that
the Hopf algebra $(P^R_\mu,M_j,N_j)$, to which we were led by
our time-to-the-right ordering convention, turns out
to be a well known example of \kkP\ Hopf algebra, the Majid-Ruegg
algebra first discussed in Refs.~\cite{MajidRuegg,lukieAnnPhys}.
Instead the $(P^S_\mu,M_j,N_j)$ algebra, which we ended up
considering starting from the time-symmetrized ordering convention,
is actually a new example of \kkP\ Hopf algebra, which had
not previously emerged in the literature.

While for a mathematical discussion of the properties of the
Majid-Ruegg $(P^R_\mu,M_j,N_j)$ algebra we refer the reader to
the literature~\cite{MajidRuegg,lukieAnnPhys}, it seems appropriate
to note here some of the key characteristics of
our $(P^S_\mu,M_j,N_j)$ Hopf algebra.
The algebra sector of $(P^S_\mu,M_j,N_j)$ is characterized by
commutation relations
\bea
\qs P_{\mu}^S,P_{\nu}^S\qd&=&0\nn\\
\qs M_j,M_k\qd&=&i\varepsilon_{jkl}M_l\;\;\qs {\mN}_j,M_k\qd=
i\varepsilon_{jkl}{\mN}_l \;\;\qs {\mN}_j,{\mN}_k\qd=-i\varepsilon_{jkl}M_l\nn\\
\qs M_j,P^S_0\qd&=&0\;\;\;[M_j,P_k^S]=i\epsilon_{jkl}P_l^S\nn\\
\qs {\mN}_j,P^S_0\qd&=&ie^{-\frac{\lambda}{2}P^S_0}P_i^S\nn\\
\qs {\mN}_j,P_k^S\qd&=&i
e^{-\frac{\lambda}{2}P^S_0}\qs\ts \frac{\sin(\lambda P^S_0)}{\lambda}
+\frac{\lambda}{2}\vec{P^S}^2\td\delta_{jk}-\frac{\lambda}{2}P_j^SP_k^S\qd\nn
~,
\eea
with ``mass" Casimir
\be
C(P^S)=\cosh(\lambda P^S_0)-\frac{\lambda^2}{2}\vec{P^S}^2
~,
\ee
while the co-algebra sector is characterized by
\bea
\Delta(P^S_0)&=&P^S_0\otimes 1+1\otimes P^S_0\;\;\;\Delta(P_i^S)
=P_i^S\otimes e^{\frac{\lambda}{2}P^S_0}
+e^{-\frac{\lambda}{2}P^S_0}\otimes P^S_j\nn\\
\Delta(M_i)&=&M_i\otimes 1+1\otimes M_i\nn\\
\Delta({\mN}_j)&=&{\mN}_j\otimes 1+e^{-\lambda P^S_0}\otimes {\mN}_j
+\lambda\epsilon_{jkl}e^{-\frac{\lambda}{2}P^S_0}P_j\otimes M_k
\eea

Just like the Majid-Ruegg $(P^R_\mu,M_j,N_j)$ algebra,
also $(P^S_\mu,M_j,N_j)$ is a bicrossproduct Hopf
algebra~\cite{MajidRuegg,lukieAnnPhys}.
Since the  Majid-Ruegg $(P^R_\mu,M_j,N_j)$ algebra
is often referred to as the ``bicrossproduct basis", it may
be natural to call ``type-2 bicrossproduct basis"
our $(P^S_\mu,M_j,N_j)$ Hopf algebra.

The reader familiar with the relevant literature will recognize
that $\Delta(P^S_0)$, $\Delta(P_i^S)$ and $C(P^S)$
for the Hopf algebra $(P^S_\mu,M_j,N_j)$ are identical to the ones
of the so-called \kkP\ ``standard basis" Hopf
algebra~\cite{LNRT}.
However, other characteristics of $(P^S_\mu,M_j,N_j)$
are different from the ones of the ``standard basis", which
in particular is not of bicrossproduct type.

\section{More on the description of translations}
The results reported in the previous Sections
were based on the concept of symmetry codified
in (\ref{twopartsnc}). This is a natural symmetry requirement, which however
deserves a few more comments, which will also lead us to a noteworthy
characterization of translations.
Our discussion of this point will be more easily followed as we consider
a specific action. Let us focus on the simple example
\be
S(\phi)=\int{\de}^4{\x}\;\phi^2 ~.
\ee
It is natural to describe infinitesimal translations in terms
of
\bea
{\x}&\rightarrow&{\x}'={\x} - \alpha \e\nn\\
\phi({\x})&\rightarrow& \phi'({\x})
=\phi({\x})+i\alpha T\phi({\x})+O(\alpha^2)\nn
\eea
where $T=-i{\e}^{\mu}\partial_{\mu}$ is the
generator of translations and $\alpha\in R$ is an expansion parameter.

On the basis of an analogy with corresponding analyses in commutative
spacetimes there are actually two possible starting points
for a description of $T$ as a symmetry of the action:
\be
\delta_{I} S(\phi)= i \int{\de}^4{\x}\;T{\cdot}\phi^2 = 0 \label{sostanza}
\ee
and
\be
\delta_{II} S(\phi)=S(\phi')-S(\phi) = 0 \label{forma}
\ee
In the context of theories in commutative spacetimes the conditions
(\ref{sostanza}) and (\ref{forma})
are easily shown to be equivalent.
But in a noncommutative space-time this is not necessarily the case.
Let us start by considering the simplest possibility, the case
of commutative transformation parameters $\epsilon$. In this
case by expanding (\ref{sostanza}) and (\ref{forma}) to
first order one finds
\begin{eqnarray*}
\delta_{II} S(\phi) =
S(\phi+i\alpha\e P\phi)-S(\phi) &=& i\alpha\int{\de}^4{\x}\Big\{({\e}
P\phi)\phi+\phi({\e} P\phi)+i\alpha({\e} P\phi)({\e} P\phi)\Big\}=
\\ &=& i\alpha\int{\de}^4{\x}\Big\{({\e} P\phi)\phi+\phi({\e}
P\phi)\Big\}+o(\alpha) \\
\delta_{I} S(\phi)= \int{\de}^4{\x}i\alpha{\e} P\phi^2 &=&
i\alpha\int{\de}^4{\x}
\Big\{{\e}_\mu(P^\mu\phi)\phi+{\e}_0\phi(P^0\phi)+{\e}_j(e^{-\lambda
P_0}\phi)(P^j\phi)\Big\} ~,
\end{eqnarray*}
which clearly indicates that the condition $\delta_{I} S(\phi)=0$
does not imply $\delta_{II} S(\phi)=0$ and {\it vice versa}.
If we want to preserve the double description (\ref{sostanza})
and (\ref{forma}) of symmetry under translation transformations
we must therefore introduce noncommuative transformation parameters.
In fact, it is easy to verify that assuming
\begin{displaymath}
[\e_0,x_\mu]=0
 ~,~~~ \Phi\e_j=\e_j (e^{-\lambda P_0}\Phi) ~,
\end{displaymath}
which follow from
\begin{displaymath}
[\e_j,\x_0]=i\lambda\e_j ~,~~~ [\e_j,\x_k]=0 ~,
\end{displaymath}
one finds that the conditions (\ref{sostanza}) and (\ref{forma}) are
equivalent. Interestingly this choice of noncommutativity of the
transformation parameters allows to describe them as differential
forms\footnote{Note that this is one of the two differential calculi
introduced in Ref.~\cite{oeckdiff}.}, $\e_\mu=\de\x_\mu$,
and therefore the condition for invariance of the action
under translation transformations can be cast in the
form
\begin{displaymath}
S(\Phi+i\de\x P\Phi)-S(\Phi)=i\de\x_\mu\int\de^4\x\;P^\mu\Phi^2
\end{displaymath}

It appears plausible that other
choices of noncommutative transformation parameters would preserve
the double description (\ref{sostanza})
and (\ref{forma}) of symmetry under translation transformations.
But it appears likely that this connection with differential
forms has some deep meaning: in order to preserve the
possibility to describe translation symmetries according to (\ref{forma})
(having already verified that these symmetries satisfied the first
criterion (\ref{sostanza}))
we ended up adopting the following description of translations
\bea
{\x}_\mu&\rightarrow&{\x}_\mu'={\x}_\mu + \de\x_\mu \nn\\
\phi({\x})&\rightarrow& \phi'({\x})
=\phi({\x})+ i\de\x_\mu P^\mu\Phi \nn
\eea
where the $\de\x_\mu$ describe the proper concept, as previously
established~\cite{oeckdiff}, of differential forms for our noncommutative
spacetime and the $P^\mu$ act as previously
described ($P_\mu\Omega_R(f)=\Omega_R(-i\partial_\mu f)$).
This is rather satisfactory from a conceptual perspective, since
even in commutative spacetime an infinitesimal translation
is most properly described as ``addition" of a differential form.

The differentials satisfy nontrivial commutation relations~\cite{oeckdiff}
\begin{displaymath}
[\de\x_0,\x_\mu]=0\quad[\de\x_j,\x_k]=0\quad [\de\x_j,\x_0]=i\lambda\de\x_j
\end{displaymath}
as required for our translations to preserve the $\kappa$-Minkowski
commutation relations (again with ${\x}_\mu'={\x}_\mu + \de\x_\mu$)
\bea
[{\x}_j,{\x}_k]=0  \rightarrow  [{\x}_j',{\x}_k']=0~,~~~~
[{\x}_j,{\x}_0]=i\lambda {\x}_j \rightarrow [{\x}_j',{\x}_0']=i\lambda {\x}_j'
\nn
\eea

An infinitesimal translation $\Phi\to \Phi' \equiv \Phi+\de\Phi$
associates to each element of $\kappa$-Minkowski (which we here
denote by $M_\kappa$) an element of the algebra $M_\kappa\oplus \Gamma$
defined over a vector space that is direct sum
of $\kappa$-Minkowski and
the bimodule~\cite{oeckdiff}, $\Gamma$, over $M_\kappa$,
with product rule
\bea
(\Phi+\de\Phi)(\Psi+\de\Psi)
=\Phi\Psi+\Phi{\cdot}\de\Psi+\de\Phi{\cdot}\Psi=\Phi\Psi+\de(\Phi\Psi)
\label{prodmod}
\eea
This algebra is isomorphic to $\kappa$-Minkowski through the map $1+\de$, which
is an algebra-isomorphism. Then an infinitesimal translation transforms an
element of $\kappa$-Minkowski in an element of a ``second copy" of
$\kappa$-Minkowski. It is a transformation internal to the
same {\underline{abstract}} algebra.
This abstract algebra {\underline{is}} our ``space of
functions of the spacetime coordinates".

The careful reader can easily verify that the equivalence of (\ref{sostanza})
and (\ref{forma}) emerged as a result of the fact that, while for our
translation generators $P_\mu$ the Leibniz rule clearly does not hold,
for the infinitesimal translation ``$\de$"
the Leibniz rule holds:
\begin{displaymath}
\de(\Phi\Psi) = (\Phi\Psi)'- \Phi\Psi = \de\Phi \, \Psi + \Phi \, \de\Psi
\end{displaymath}
(where again  $\Phi' \equiv \Phi+\de\Phi$ and we used
the product rule (\ref{prodmod})).

This network of results and interpretations provides the conceptual
ground for a description of translational symmetry
in $\kappa$-Minkowski spacetime.
Simply by inspection of the $\kappa$-Minkowski commutation
relations one already concludes that classical translations
cannot be a symmetry of this spacetime.
The question then is whether translations are a lost/broken symmetry
of this spacetime or instead they are simply deformed by
the $\kappa$-Minkowski noncommutativity.
We have shown that one can construct theories in $\kappa$-Minkowski
spacetime that enjoy a deformed/quantum (Hopf-algebra) translational
symmetry.

Again by inspection of the $\kappa$-Minkowski commutation
relations one can see that instead classical rotations can
be implemented as a symmetry. But for boosts something analogous
to what happens for translations occurs: classical boosts are not
a symmetry of $\kappa$-Minkowski, but, as we showed, there is
a quantum/deformed version of boosts that are symmetries.

\section{Replacing the classical Poincar\'{e} algebra}
One other point which must be stressed within our analysis
is that both the algebra $(P^R_\mu,M_j,{\mN}_j)$ and
the algebra $(P^S_\mu,M_j,{\mN}_j)$
involve underformed Lorentz sector, meaning that the commutators
in the rotation/boost sector are unmodified.
However, the action of boosts on momenta is modified in
a nonlinear way. The action of boosts on momenta represents
a nonlinear realization of the Lorentz group.
These are the ingredients of the most popular realizations
of the ``DSR" (or ``Doubly Special Relativity")
framework~\cite{dsr1,Kow02-NST,Mag01-DSR}.
The DSR nonlinearities are the way in which
a second relativistic invariant (in addition to speed-of-light scale)
becomes manifest. The second invariant is, in the notation here adopted,
the length scale $\lambda$, which can be naturally identified
with the Planck length.

The fact that the Lorentz group is still present, but is nonlinearly
realized in some DSR proposals, has led some authors~\cite{ahluOPERAT,grumi}
to argue that essentially the nonlinearity might be a formal artifact,
that one should unravel the nonlinearity by performing some nonlinear
redefinition of the energy-momentum variables, and that ``true" translations
would still have to satisfy linear commutation relations with boosts.

The analysis we are reporting exposes a hidden assumption of the arguments
advocated in Refs.~\cite{ahluOPERAT,grumi}: the authors are clearly
assuming that the underlying spacetime be classical (or at least
commutative).
We showed explicitly that in some examples of spacetime noncommutativity
translations are inevitably a nontrivial concept.
In both our examples, the algebra $(P^R_\mu,M_j,{\mN}_j)$ and
the algebra $(P^S_\mu,M_j,{\mN}_j)$, boosts act nonlinearly on the one-particle
sector (which is faithfully described by the algebra sector)
and on the multi-particle sector (which in these noncommutative spacetimes
is described through the coalgebra sector of the Hopf algebra of symmetries).
We have only constructed two illustrative examples of symmetry algebras,
but the careful reader can easily verify that no change of ordering
convention for functions of the \kM\ coordinates can fully remove
the nonlinearity. Some form of nonlinearity will inevitably
survive\footnote{At the cost of an awkward nonlinear redefinition
of energy-momentum~\cite{judesvisser} (which would not correspond
to a workable ordering conventions in \kM\ and would not have a natural
associated differential calculus), one can even remove $\lambda$
completely from the algebra sector and could be misled into believing that
the scale $\lambda$ would have been ``redefined away". But instead
the same redefinition of energy-momentum that trivializes the algebra
sector requires, in order to define coherently the action of
these new tiem-space translations generators on products of functions,
that the coproduct be  $\lambda$-dependent in a highly nontrivial way.
For this awkward setup the action of boosts on one-particle states
would still be linear, but only at the cost of a highly nonlinear action
on two-particle states.},
since this is the natural energy-momentum manifestation of the
fact that the \kM\ commutation relations, with their length scale $\lambda$,
are being introduced as a fundamental feature of spacetime.
In \kM\ the description of translations, as shown here,
necessarily requires some new structures, as
most elementarily seen by looking
at the product of plane waves in $\kappa$-Minkowski
noncommutative spacetime which we already discussed above:
$[e^{- i k x} e^{i E t}] {\cdot} [e^{- i q x} e^{i \omega t}]
=[e^{- i (k + e^{\lambda E} q) x} e^{i (E+\omega) t}]$.
Clearly the spacetime noncommutativity is leading to a law of composition
of energy
momentum $((k,E),(q,\omega)) \rightarrow ((k + e^{\lambda E} q),(E+\omega))$
which is not covariant under ordinary boosts and involves a clear nonlinearity.
In this type of contexts a nonlinear realization of the Lorentz group
is the only way to save the equivalence of inertial frames
(the ``Relativity Principle").

And there starts to be growing evidence~\cite{kodadsr}
that noncommutativity is not the only realization of the idea
of ``spacetime quantization" that leads to DSR-type implementation
of the Relativity Principle. In particular, in Ref.~\cite{kodadsr}
it was recently shown that in a large class of quantum-gravity
theories a quantum (Hopf-algebra), rather than classical, symmetry
is realized when the cosmological constant is nonzero. And
the  symmetries are still quantum rather than classical
in the limit of vanishing cosmological constant
(the limit in which these quantum symmetries, of
course, are Poincar\'{e}-like).
Again the nonlinearities of the boost action turn out
to be unavoidable: one appears to have some freedom to transfer
some of the nonlinearity form the one-particle to the two- (multi-)particle
sector or viceversa, but the presence of some nonlinearity is
unavoidable.

The DSR framework was introduced~\cite{dsr1} as a contribution
to quantum-gravity research and therefore it implicitly assumes
that spacetime has strikingly nonclassical features, such as noncommutativity.
By focusing, as essentially done in Ref.~\cite{ahluOPERAT,grumi},
on the inadequacy of the DSR framework
for theories in a classical spacetime one is basically missing the
key objective of the DSR proposal. DSR was never
intended to apply in a classical spacetime. And actually
the point should be properly phrased in reverse: the DSR frameworks
which we are here considering, the ones with nonlinear realization
of the Lorentz group on energy-momentum space, take energy-momentum
space as primitive entity (in the sense of operative definitions)
and spacetime as a derived entity\footnote{Thinking in these terms
may be challenging at first (or at least it was at first challenging
for us), since our present world view implicitly
assumes that spacetime entities are primary. For example,
we describe energy-momentum
through the concept of translations in spacetime. It is hardly ever
stressed that spacetime
coordinates may be similarly described through
translations in energy-momentum space.},
and therefore one should not ask in which way can the nonlinearity
in energy-momentum space be unravelled (there will always be a trivial
answer: it can be unravelled by assuming it refers to a classical spacetime)
but rather one should look for a spacetime picture in which the nonlinearity
plays a natural role. The type of noncommutative spacetimes here considered
is an example of natural application of the DSR concepts.

\section{Summary and outlook}
The analysis we reported in the previous Sections has provided insight
on some questions, and raised new questions, for the study of
physical theories in certain types of noncommutative spacetimes.
We have introduced a concept of noncommutative-spacetime symmetry,
which follows very closely the one adopted in commutative spacetimes,
and actually is analyzed most naturally in terms of a Weyl map,
{\it i.e.} with the support of familiar properties of commutative
functions in an auxiliary  commutative spacetime.
We argued that the symmetries of a theory in such a noncommutative
spacetime may depend rather strikingly on the specific form of
the action that characterizes the theory. While theories in
classical Minkowski spacetime are always
Poincar\'{e} invariant, if invariant under a 10-generator algebra,
it appears plausible that different theories in \kM\ spacetime
might enjoy different 10-generator symmetries (different Hopf-algebra
versions of the Poincar\'{e} symmetries).

We found that for a specific simple theory, a theory describing a
free scalar field in \kM, it was impossible to find a formulation
that would admit invariance under ordinary
(classical) Poincar\'{e} transformations.
This, in particular, allowed us (Section VIII)
to clarify a concern about DSR theories
which had been raised in  Ref.~\cite{ahluOPERAT,grumi}, by showing that
those studies had implicitly relied on the assumption that the spacetime
would be commutative and essentially classical.
More importantly for the key objectives of our analysis,
we did find (Section IV) 10-generators symmetries of our description
of a free scalar field in \kM, and these symmetries admitted formulation
in terms of Hopf-algebra (quantum)
versions of the classical Poincar\'{e} symmetries.

Although our analysis allowed us to reduce the amount of ambiguity
in the description of the symmetries of theories in these noncommutative
spacetimes, we are left with a choice between different
realizations of the concept of translations in the noncommutative spacetime.
We have clarified that such an ambiguity might have to be expected on the basis
of the type of coordinate noncommutativity here considered, but it remains
to be seen whether by appropriate choice of the action of the theory
one can remove the ambiguity, {\it i.e.} construct a theory which is
invariant under one specific type of translations and not
under any other type.
This appears to be the most urgent open issue that emerges from
our study, and, as argued in Section~V, one natural context
in which to explore this issue might be provided by attempting
to construct gauge theories in \kM\ spacetime following the
approach here advocated.

\section*{Acknowledgments}
We are grateful for conversations G.~Mandanici during the preparation
of this manuscript.
We also thank J.~Kowalski-Glikman and E.~Witten: conversations with
them provided encouragement for us to rewrite more pedagogically
Section~VII (which offered fewer details in an earlier version of the
manuscript).

\appendix\section{Deformed boost generators}\noindent
As announced in Subsection~IV.E, in this
appendix we report some aspects of the analysis necessary
in order to construct the boost generators ${\mathcal{N}}^R$.
Analogous techniques can be used for ${\mathcal{N}}^S$,
but here we focus on ${\mathcal{N}}^R$, and, since
we therefore always refer to the ``right-ordered" $\Omega_R$ map,
the label $R$ is omitted.

The starting point for the analysis reported in Subsection~IV.E
was the 7-generator Hopf algebra with $P_{\mu}$
and $M_j$ generators, which we note again here for convenience:
\bea
[P_\mu,P_\nu]=0 ~, ~~~~~ [M_j,P_0]=0, &\;\;&
 ,~~~~~ [M_j,P_l]=i\varepsilon_{jlm}P_m
 \nn\\
P_{\mu}\Omega(e^{ikx})=\Omega(-i\partial_{\mu}e^{ikx})&\;\;&
 ,~~~~~ \Delta P_{\mu}
=P_{\mu}\otimes 1+e^{\lambda P_0(\delta_{\mu 0}-1)}\otimes P_{\mu}\nn\\
M_j\Omega(e^{ikx})=\Omega(-i\epsilon_{jkl}x_k\partial_le^{ikx})&&
  ,~~~~~ \Delta M_j
=M_j\otimes 1+1\otimes M_j\nn
\eea
As showed in Subsection~IV.D, one cannot extend this 7-generator algebra
to a 10-generator algebra by adding ordinary (classical) boost generators.
But one can obtain a 10-generator Hopf algebra by introducing deformed
boost generators ${\mathcal{N}}$, and in particular it is possible
to do so while leaving the Lorentz-sector commutation relations
unmodified
\bea
&&[M_j,M_k]=i\varepsilon_{jkl}M_l\nn\\
&&\qs {\mathcal{N}}_j,M_k\qd=i\varepsilon_{jkl}{\mathcal{N}}_l\nn\\
&&\qs {\mathcal{N}}_j,{\mathcal{N}}_k\qd=-i\varepsilon_{jkl}M_l ~.
\label{com:NN}
\eea
We intend to show this explicitly here.

Let us start observing that
the most general form in which the commutation
relations among ${\mathcal{N}}_j$
and $P_l$ can be deformed (consistently with the underlying space-rotation
symmetries) is
\bea
&&[{\mathcal{N}}_j,P_0]=iA(P)P_j\nn\\
&&\qs {\mathcal{N}}_j,P_l\qd=i\lambda^{-1}B(P)\delta_{jl}+i\lambda C(P)P_jP_l
+iD(P)\varepsilon_{jlm}P_m\label{com:PN}
\eea
where $A,B,C,D$ are unknown adimensional
functions of $\lambda P_0$ and $\lambda^2\vec{P}^2$.
In the classical limit $A(P)=1,\;\;\lambda^{-1}B(P)=P_0,\;\;D=0$
and $C$ can take any form, as long as it is finite, since $\lambda C(P)$
must vanish in the classical $\lambda \rightarrow 0$ limit.

From these relations it is easy to find that ${\mathcal{N}}_j$ can be
represented as differential operators inside the $\Omega$ map,
\bea
{\mathcal{N}}_j \Omega(\phi(x))=\Omega\{[ix_0A(-i\partial_x)\partial_j
+\lambda^{-1}x_jB(-i\partial_x)-\lambda x_lC(-i\partial_x)\partial_l\partial_j
-i\epsilon_{jkl} x_k D(-i\partial_x)\partial_{l}]\phi(x)\}~, \nn
\eea
and using $\Omega_R(x_0f)=[{\x}_0-\lambda\vec{\x}\vec{P}]\Omega_R(f)$
on can rewrite ${\mathcal{N}}_j$ as
\begin{displaymath}
{\mathcal{N}}_j=\Big(\lambda^{-1}{\x}_jZ(\ar)-{\x}_0P_j\Big)A(\ar)-\lambda
C(\ar)\e_{jkl}P_kM_l+D(\ar)M_j
\end{displaymath}
where $Z \equiv B-\lambda^2\vec{P}^2C$.

In preparation for the rest of the analysis, we note here that
\begin{eqnarray*}
&& {\x}_j\Omega_R(f)=\Omega_R(x_jf)=\Big(e^{\lambda P_0}\Omega_R(f)\Big){\x}_j
\end{eqnarray*}
and we introduce the useful notations $g_p=e^{ip_1{\x}_1}e^{-ip_0{\x}_0}$
and
$p \cplus q=(p_0+q_0,p_1+q_1e^{-\lambda p_0},0,0)$,
so that $g_p g_q=g_{(p \cplus q)}$. For a
generic scalar function it will be implictly assumed that it depends
on the operators $P_0$ and $P^2$ (so for example, with $A$ we will
denote $A(P_0,P^2)$). A notation of type $A(q_0,q^2)$
will be reserved to functions which depend on a real four-vector $q$.

Since
\begin{displaymath}
M_1g_p=P_2g_p=P_3g_p=0
\end{displaymath}
it is easy to verify that
\begin{displaymath}
{\mathcal{N}}_1\,g_{(p\cplus q)}=A(p\cplus q)\Big(\lambda^{-1}Z(p\cplus
q){\x}_1-{\x}_0(p_1+q_1e^{-\lambda p_0})\Big)\,g_{(p\cplus q)}
\end{displaymath}
and we can remove ${\x}_0$ from the previous expression using the identity
\begin{displaymath}
{\x}_0P_1\,g_p=(\lambda^{-1}{\x}_1Z-{\mathcal{N}}_1\,A^{-1})g_p
\end{displaymath}
obtaining
\begin{eqnarray*}
{\mathcal{N}}_1\,g_{(p\cplus q)} &=& ({\mathcal{N}}_1
A^{-1}A_{(1)}g_p)(A_{(2)}g_q)+(e^{-\lambda
P_0}A_{(1)}g_p)({\mathcal{N}}_1 A^{-1}A_{(2)}g_q)+ \\ &&
+\lambda^{-1}{\x}_1
Ag_pg_q\left\{Z(p\cplus q)-Z(p)-e^{-2\lambda p_0}Z(q)
-\lambda^2e^{-\lambda p_0}p_1q_1\right\}
\end{eqnarray*}
We notice that it is impossible to eliminate ${\x}_1$
from this expression without reintroducing ${\x}_0$,
and therefore, if we
want the second member to be function
only of ${\mathcal{N}}_j$ (and of $P_0$, $P_1$ and
operators with null action on $g_p$),
the factor which multiplies ${\x}_1$ must
vanish identically. For this, it is necessary and sufficient that
\begin{equation}\label{eq:cond}
Z\Big(p_0+q_0,(p_1+q_1e^{-\lambda p_0})^2\Big)=Z(p_0,p_1^2)+e^{-2\lambda
p_0}Z(q_0,q_1^2)+\lambda^2e^{-\lambda p_0}p_1q_1
\end{equation}
for each $p$ and $q$.

We must now find a solution of (\ref{eq:cond}).
Applying $\partial_{p_1}\partial_{q_1}$ to (\ref{eq:cond}) we obtain
\begin{displaymath}
\partial_{p_1}\partial_{q_1} Z\Big(p_0+q_0,(p_1
+q_1e^{-\lambda p_0})^2\Big)
=\lambda^2e^{-\lambda p_0}
\end{displaymath}
In the first member we can substitute $\partial_{q_1}$
with $e^{-\lambda p_0}\partial_{p_1}$, and
calculate the resulting expression in $q_0=q_1=0$,
leading us to
\begin{displaymath}
\partial_{p_1}^2Z(p_0,p_1^2)=\lambda^2
\end{displaymath}
This can be integrated to
\begin{displaymath}
Z(p_0,p_1^2)=W(\lambda p_0)+\beta\lambda p_1+\lambda^2p_1^2/2
\end{displaymath}
with $\beta$ and $W$ arbitrary.

Since $Z$ is function only of $p_0$ and $p_1^2$, $\beta$
must be zero. Substituting in (\ref{eq:cond}) we get the condition
\begin{displaymath}
W(\lambda p_0+\lambda q_0)=W(p_0)+e^{-2\lambda p_0}W(\lambda q_0)
\end{displaymath}
For $p_0=q_0=0$ we have $W(0)=2W(0)$, that is $W(0)=0$.

Applying
$\partial_{\lambda q_0}$, using the identity $\partial_{\lambda q_0}W(\lambda
p_0+\lambda q_0)=\partial_{\lambda p_0}W(\lambda p_0+\lambda q_0)$ and
calculating in $q_0=0$ we obtain
\begin{displaymath}
\partial_{\lambda p_0}W(\lambda p_0)
=e^{-2\lambda p_0}\partial_{\lambda q_0}W(\lambda q_0)\big|_{q_0=0}
\end{displaymath}
which integrated, with the condition $W(0)=0$, gives
\begin{displaymath}
W(\lambda p_0)=\frac{1-e^{-2\lambda p_0}}{2}\,\partial_{\lambda p_0}W(\lambda
p_0)\big|_{p_0=0} ~.
\end{displaymath}

From the request that ${\mathcal{N}}_1$ has the correct classical limit,
 we derive $\partial_{\lambda p_0}W(0)=1$, and this allows us to
 write
 \begin{displaymath}
Z(p_0,p_1^2)=\frac{1-e^{-2\lambda p_0}}{2}+\frac{\lambda^2}{2}p_1^2 ~.
\end{displaymath}

Since $Z$ is a scalar, the knowledge of $Z(p_0,p_1^2)$ is equivalent to the
knowledge of $Z(p_0,p^2)$ for a generic $\vec{p}$
(just renaming $p_1^2\to p^2$).
We have therefore established that
\begin{displaymath}
{\mathcal{N}}_j=\tilde{{\mathcal{N}}}_jA-i\lambda C\e_{jkl}P_kM_l+DM_j
\end{displaymath}
where $A,C,D$ are arbitrary adimensional scalar function of $P_\mu$ and
\begin{displaymath}
\tilde{{\mathcal{N}}}_j={\x}_j\left\{\frac{1-e^{-2\lambda
P_0}}{2\lambda}+\frac{\lambda}{2}P^2\right\}-{\x}_0 P_j
\end{displaymath}

It is straightforward
to verify that the 10 generators $({\mathcal{N}},M,P)$ close a Hopf algebra,
for any choice of
the triplet $(A,C,D)$. In fact, the coproduct of ${\mathcal{N}}_j$ is
\begin{displaymath}
\Delta {\mathcal{N}}_j=\left({\mathcal{N}}_j\otimes 1+e^{-\lambda P_0}\otimes
{\mathcal{N}}_j+\lambda\e_{jkl}P_k\otimes M_l\right)\Delta A
+\lambda \e_{jkl}(\Delta C)(\Delta
P_k)(\Delta M_l)+\Delta D{\cdot}\Delta M_j
\end{displaymath}
where $(\Delta A,\Delta C,\Delta D)$ are known tensors,
for each choice of $(A,C,D)$.

The request of a classical Lorentz subalgebra ($M_j$,$\tilde{{\mathcal{N}}}_j$)
leads to the conditions
\be
D=0\quad\qquad [\lambda A\partial_0
-2(C+\lambda^2 P^2D)\partial_{\vec{P}^2}+\lambda^2 D] C=-\lambda^2
\label{condjoc}
\ee
The deformed boost operators considered in
Subsection~IV.D correspond to the $A=1,C=D=0$ solution of (\ref{condjoc})
({\it i.e.} correspond to $\tilde{{\mathcal{N}}}_j$)
and take the form
\begin{displaymath}
{\mathcal{N}}_j\Omega_R(f)=\Omega_R\ts \qs ix_0\partial_j
+x_j\ts\frac{1-e^{2i\lambda \partial_0}}{2\lambda}
-\frac{\lambda}{2}\nabla^2\td
+\lambda x_l\partial_l\partial_j\qd f\td ~.
\end{displaymath}





\bigskip



\end{document}